\documentclass[useAMS,usenatbib,referee]{biom}
\def\bSig\mathbf{\tau}

\usepackage[figuresright]{rotating}
\usepackage{multicol, multirow}
\usepackage{amsmath}
\usepackage{natbib}
\usepackage{hyperref}
\usepackage{bigints}
\usepackage{xcolor}

\DeclareUnicodeCharacter{2061}{}
\DeclareUnicodeCharacter{FF0C}{}

\title[Sensitivity analysis in meta-analysis of sparse data]{Sensitivity analysis for publication bias in meta-analysis of sparse data based on exact likelihood}

\author{Taojun Hu\\
	 Department of Biomedical Statistics, Graduate School of Medicine, Osaka University, Osaka, Japan\\
 Department of Biostatistics, School of Public Health, Peking University, Beijing, China
	 \and 
	 Yi Zhou\\
	 Beijing International Center for Mathematical Research, Peking University, Beijing, China
 \and
 Satoshi Hattori$^{*}$\email{hattoris@biostat.med.osaka-u.ac.jp}\\
	 Department of Biomedical Statistics, Graduate School of Medicine, Osaka University, Osaka, Japan\\
 Integrated Frontier Research for Medical Science Division, \\Institute for Open and Transdisciplinary Research Initiatives (OTRI), Osaka University, Osaka, Japan\\
 }

\begin{document}





\label{firstpage}


\begin{abstract}
Meta-analysis is a powerful tool to synthesize findings from multiple studies. The normal-normal random-effects model is widely used to account for between-study heterogeneity. However, meta-analysis of sparse data, which may arise when the event rate is low for binary or count outcomes, poses a challenge to the normal-normal random-effects model in the accuracy and stability in inference since the normal approximation in the within-study model may not be good. To reduce bias arising from data sparsity, the generalized linear mixed model can be used by replacing the approximate normal within-study model with an exact model. Publication bias is one of the most serious threats in meta-analysis. Several \textcolor{blue}{quantitative} sensitivity analysis methods for evaluating the potential impacts of selective publication are available for the normal-normal random-effects model. We propose a sensitivity analysis method by extending the likelihood-based sensitivity analysis with the $t$-statistic selection function of Copas to several generalized linear mixed-effects models. Through applications of our proposed method to several real-world meta-analysis and simulation studies, the proposed method was proven to outperform the likelihood-based sensitivity analysis based on the normal-normal model. The proposed method would give useful guidance to address publication bias in meta-analysis of sparse data.
\end{abstract}

%

\begin{keywords}
Generalized linear mixed model; Meta-analysis of sparse data; Publication bias; Research synthesis; Selection function; Sensitivity analysis
\end{keywords}


\maketitle

\section{Introduction}
\label{ss:intro}

Meta-analysis is a technique to integrate results of multiple studies and then establish sound evidence of higher precision. It is widely used in medical, biological, and pharmaceutical research and then often plays important roles
in data-oriented decision-making. \textcolor{blue}{Suppose we are interested in evaluating the effect of an experimental treatment by synthesizing multiple studies. In this paper, we consider meta-analysis based on information extracted from literatures. In words, the summary statistics representing effectiveness of the treatment such as the log odds ratio of each individual study are used as the outcome.} The conclusions of each study drawn by these summary statistics might not be sufficiently precise, and might be qualitatively
heterogeneous. Even if qualitatively consistent, the treatment effects might vary substantially since their
protocols were not necessarily consistent and the patients involved might have heterogeneous characteristics. In these cases,
meta-analysis would provide us with very informative insights into the results by synthesizing these various studies to obtain
a more comprehensive conclusion. In aggregating the results of the multiple studies in meta-analysis, the fixed-effect
and the random-effects models are the most widely applied methods. The fixed-effect model assumes between-study
homogeneity, that is, all studies share a common mean effect size.
On the other hand, the random-effects model allows us to incorporate heterogeneous effect sizes over studies by using random-effects. It has two-level structure to decompose the variations into two parts: the variation within individual study and that across studies. The former is due to errors in estimation and the latter is modeled with the random-effects. In this paper, the model describing the estimation error within individual study is called the within-study model and the corresponding likelihood the within-study likelihood. The model for the random-effects to describe the across-study variation is called the across-study model. With this flexibility, the random-effects model is more appealing and informative since the
assumption of homogeneity may not be realistic in most meta-analysis. A general review of the random-effects meta-analysis was given by \cite{stijnen2010}. The most widely used random-effects model would be the normal-normal (NN) model, which relies on the asymptotic normal approximation in the within-study model.

Although the NN model provides a simple way for synthesizing evidence, the validity of the normal approximation in modeling the within-study variation may be criticized since
it may be unrealistic in sparse data cases; for studies with limited numbers of subjects or events, the
asymptotic approximation based on the central limit theory may be questionable. \textcolor{blue}{To address the problem in meta-analysis of sparse data, \cite{cai2010meta} proposed a Poisson-Gamma hierarchical model based on the Poisson random-effects model. 
\cite{bohning2015meta} and \cite{kuss2015statistical}
pointed out that one can handle meta-analysis of zero events without relying on continuity correction and proposed the Poisson-normal hierarchical model and the beta-binomial regression model to this end, respectively. \cite{stijnen2010} proposed to use the generalized linear mixed models (GLMMs) to replace the approximate normal distribution for the within-study model with the exact distribution.} 
Since the GLMMs had substantial advantages in meta-analysis of sparse data, we focus on the GLMMs.

Scientific papers of individual studies are more likely to be published by
scientific journals with striking results. Thus, if meta-analysis was conducted only on papers published in scientific
journals, one may miss negative results and then the aggregated estimates in meta-analysis may be biased. This bias is referred to as publication bias (PB) \citep{begg1988,dickersin1990,easterbrook1991}. The issue of PB has raised much attention from researchers from an
early age and statistical methods against the PB have been extensively studied. \textcolor{blue}{Graphical methods
such as the funnel-plot and the trim-and-fill methods proposed by \cite{egger1997bias}, \cite{duval2000trim} and \cite{duval2000nonparametric}} provided intuitively understandable methods to examine whether the
meta-analysis result struggled with PB. However, insights given by these graphical methods are not based on sound modeling of the potential
selective publication process and then may be subjective and misleading. More \textcolor{blue}{quantitative} methods have been
successfully introduced based on selection functions to address the impacts of PB less subjectively. 

\cite{copas1999} and \cite{copas2000,copas2001} introduced the Heckman type selection function, which was first introduced in the
econometric field by \cite{heckman1976} and \cite{heckman1979}, to address PB. We refer to this selection function as the Copas-Heckman selection
function. It relied on the NN model. The joint distribution of the NN model and the latent Gaussian model for the selective publication can be handled within a rather simple bivariate normal distribution. Indeed, the conditional distribution of the outcome can be derived under the bivariate normal distribution theory. On the other hand, since only outcomes of published studies can be used for inference, unidentifiability issue arises. Then \cite{copas1999} and \cite{copas2000,copas2001} took a sensitivity analysis approach fixing some unknown parameters as sensitivity parameters. Since the true sensitivity parameters are unknown, results with a set of parameters should be addressed. To avoid this rather messy process and then directly obtain bias-adjusted estimates, \cite{ning2017} proposed an EM-algorithm-based method with additional
assumptions for unpublished studies based on funnel-plot symmetry. \cite{huang2021} proposed to maximize the full likelihood, instead of the conditional likelihood given published, using information on unpublished studies from the clinical trial registries. \textcolor{blue}{On the other hand, the selective publication process defined by the Copas-Heckman selection function is not necessarily appealing. In medical research, testing hypothesis is often crucial role in scientific discussion in research papers and the $p$-value or the corresponding test statistic would play a key role in publication. A natural selective publication process may be given by a monotone function of the $p$-value or the test statistic. The Copas-Heckman selection function does not have this property. \cite{copas2013} introduced an alternative inference method to evaluate the conditional likelihood given published and successfully constructed a sensitivity analysis framework to handle the selection function monotone with respect to the $p$-value of the individual study, namely the Copas $t$-statistic selection model. Compared with the Copas-Heckman selection model, the Copas $t$-statistic selection model is easier to implement and interprete \citep{jin2015statistical}. Based on the Copas $t$-statistic selection model, \cite{huang2023} introduced an inverse probability weighting method to obtain adjusted estimates using the clinical trial registry data.}

Note that all the above methods are for the NN model. Thus, they are expected to inherit the drawbacks in the NN model for sparse data. Therefore, it is very important to develop methods to address
PB in a more general framework under the GLMMs using the exact within-study model. To the best of our knowledge, there is no method available to address publication
bias using selection functions in GLMMs. In this work, we propose a sensitivity
analysis method using the exact within-study model based on the Copas $t$-statistic selection model. We consider a
series of GLMMs and incorporate the $t$-statistic selection function into sensitivity analysis. The rest of this article is organized as follows. In Section 2, we begin by introducing our method for the simplest case of the one-group
binomial normal (BN) model. In Section 3, we present the proposed methods in more
general settings, such as the two-group binomial normal model and the hypergeometric normal (HN) model. In Section 4, we illustrate our proposal with several
real examples that have been used in previous studies. Besides the real-data
analysis, we evaluate the performance of our proposal via simulation studies in Section 5. Finally, we end our article with discussions and conclusions in Section 6. 

\section{Sensitivity-analysis for publication bias in meta-analysis of proportion based on one-group binomial normal model}
\label{ss:propose}

\subsection{Notations and an example}
\label{ss:motiv}

\textcolor{blue}{Throughout the paper, we consider literature-based meta-analysis, which rely on summary statistic and its standard error reported in scientific journals as data. We suppose that the summary statistic of each study was calculated based on individual participant data, which was independently and identically distributed sample from a binary distribution.} Suppose that we are interested in evaluating the prevalence of a disease; several studies report the
number of total subjects included in the study and that with disease. Correspondingly, we
denote those for the $i(i=1, \cdots, N)$-th study by $n_i$ and $y_i$, respectively. \textcolor{blue}{In this case, we are interested in the log odds of the prevalence of the disease, that is, the logit-transformed prevalence. The empirical log odds in the $i$-th study can be estimated by $\widehat{\theta}_i=\log (y_i /(n_i-y_i ))$; and its empirical standard deviation can be estimated by $s_i=\sqrt{1/y_i+ 1/(n_i - y_i)}$.} 

\textcolor{blue}{As an example, \cite{ngoma2019seroprevalence} conducted meta-analysis of the seroprevalence of human T-lymphotropic virus (HTLV) in blood donors in sub-Saharan Africa. We showed the funnel-plot of the log odds in Web Figure 2. The funnel-plot suggested severe PB. Studies with relatively large prevalence were less likely to be published. Thus, it is important to account for PB in the meta-analysis.}

\subsection{Copas $t$-statistics sensitivity analysis}
\label{ss:nn}

Copas $t$-statistic sensitivity analysis proposed by \cite{copas2013} is based on the NN model. The NN model assumes empirical log odds in each study follow a two-level normal model. First, the within-study level of the NN model assumes $\widehat{\theta}_i \sim N(\theta_i, s_i^2)$, where $\theta_i$ is the true log odds for the $i$-th study. Then, the heterogeneity of $\theta_i$ across studies is modeled with the random-effects, that is $\theta_i \sim N(\theta, \tau^2)$, where $\theta$ is the common log odds across studies and $\tau$ is the heterogeneity parameter suggesting the variation of log odds across individual studies in a meta-analysis. This is the across-study model for the NN model. Marginally, $\widehat{\theta}_i \sim N(\theta, s_i^2+\tau^2)$. The common log odds $\theta$ is of interest and to be estimated.

\cite{copas2013} re-expressed $(\widehat{\theta}_i, s_i)$ using $x_i=1/s_i$ and $y_i=\widehat{\theta}_i/s_i$. Consequently, the NN model translates to the conditional distribution of $y_i$ given $x_i$ as $y_i\mid x_i \sim N(\theta x_i, 1 + \tau^2 x_i^2)$, yielding the conditional density as
\begin{equation*}
 f_P(y_i\mid x_i) = \frac{1}{\sqrt{1+\tau^2 x_i^2}} \varphi\left(\frac{y_i-\theta x_i}{\sqrt{1+\tau^2 x_i^2}}\right),
\end{equation*}
where $\varphi(\cdot)$ denotes the probability density function of the standard normal distribution. \cite{copas2013} introduced a selection function to mitigate PB:
\begin{equation}
 \mathrm{P}(\text{select}\mid x_i, y_i)=a(y_i)=a(\widehat{\theta}_i/s_i),
 \label{nn.1}
\end{equation}
where $a(y_i)$ is a non-decreasing function valued from 0 to 1. This function can be arbitrary but \cite{copas2013} suggests the 2-parameter probit function, that is $a(y_i) = \Phi(\alpha+\beta y_i)$ where $\Phi(\cdot)$ denotes the cumulative distribution function of the standard normal distribution, also known as the probit function. This function determines a selection probability conditional on $x_i$, denoted as $b(x_i)$:
\begin{equation*}
 b(x_i)=\mathrm{P}(\text{select} \mid x_i)=E_{P}\{a(y_i) \mid x_i\}=\int a(y_i) f_{P}(y_i \mid x_i) dy.
\end{equation*}

The marginal selection probability $p$ is given by:
\begin{equation}
 p=\mathrm{P}(\text{select})=E_{P}\{b(x_i)\}=\int b(x_i) f_{P}(x_i) dx,
 \label{nn.2}
\end{equation}
where $f_{P}(x_i)$ is $x_i$'s population density function. Consequently, the density function for $x_i$, conditional on published, is derived as:
\begin{equation}
f_{O}(x_i) = \mathrm{P}(x_i \mid \text{select}) = \frac{b(x_i) f_{P}(x_i)}{p},
\end{equation}
and equivalently, 
\begin{equation}
  f_{P}(x_i) = \dfrac{p}{b(x_i)} f_{O}(x_i),
\label{nn.3}
\end{equation}
where the population density is denoted with suffix $P$ and the density and expectation conditional on published are denoted with suffix $O$. \textcolor{blue}{By integrating both sides of~\eqref{nn.3} over $x_i$, it holds that
\begin{equation*}
 1/p=E_O\{b(x_i)^{-1}\}.
\end{equation*}
Thus, it approximately holds that:
\begin{equation}
\frac{1}{p} = \frac{1}{N}\sum_{i=1}^N \frac{1}{b(x_i)}.
\label{nn.4}
\end{equation}}
The joint density of $(x_i, y_i)$ conditional on published is derived as,
\begin{equation}
f_{O}(x_i, y_i)=\frac{f_{P}(y_i \mid x_i) f_{O}(x_i) a(y_i)}{b(x_i)}.
\label{nn.5}
\end{equation}
Thus, the log-likelihood conditional on published is derived as:
\begin{equation}
\begin{aligned}
\ell_O(\theta, \tau) = & \sum \log \{f_{O}(x_i, y_i)\} \\
= & -\frac{1}{2} \sum \log (1+\tau^2 x_i^2) -\frac{1}{2} \sum \frac{(y_i-\theta x_i)^2}{1+\tau^2 x_i^2} +\sum \log \{a(y_i)\} -\sum \log \{b(x_i)\} \\
& +\sum \log \{f_{O}(x_i)\}.
\end{aligned}
\label{nn.6}
\end{equation}
\textcolor{blue}{where $a(y_i)$ refers to~\eqref{nn.1} and $b(x_i)$ refers to~\eqref{nn.2}. The $f_O(x_i)$ is free from the parameters.} \cite{copas2013} proposed to estimate the parameters $(\theta, \tau)$ by maximizing this conditional log-likelihood given published~\eqref{nn.6} with the constraint~\eqref{nn.4} given various marginal selection probability $p$.

\subsection{The proposed sensitivity analysis method with the binomial model}
\label{ss:onegroupbn}

The one-group binomial normal (BN) model using the exact within-study model serves as an alternative for meta-analysis of the NN model, especially in sparse data. The within-study level of the BN model adopts the binomial distribution, i.e.
\begin{equation}
y_i \mid \theta_i \sim \mathrm{Binomial}⁡\left(n_i,\frac{\exp ⁡(\theta _i)}{1+\exp ⁡(\theta _i)}\right)
\label{eq.7}
\end{equation}

As the across-study model, we suppose $\theta _i \sim N(\theta, \tau ^2)$. The population probability mass function of the BN model without considering PB can be depicted as follows  \citep{stijnen2010}, 

\begin{equation}
f_P(y_i\mid n_i) = \int \tbinom{n_i}{y_i} \scalebox{1}{$ \frac{\exp(\theta _i)^{y_i}}{\{1+\exp(\theta_i)\}^{n_i}}\frac 1{\tau }\varphi \left(\frac{\theta_i-\theta }{\tau }\right)d\theta _i$},
\label{eq.8}
\end{equation}
and the corresponding likelihood function is given by
\begin{equation}
  L(\theta ,\tau )=\prod _{i=1}^N f_P(y_i \mid n_i).
  \label{eq.add3}
\end{equation} 
One can obtain an aggregated estimate for $(\theta, \tau)$ by maximizing the likelihood $L(\theta, \tau)$~\eqref{eq.add3}. 

Taking PB into consideration, we assume
the publication probability of a study is a monotonous function of the $t$-statistics. In the one-group BN model, the $t$-statistics can be chosen as
\begin{equation*}
    t_i=\frac{\widehat{\theta}_i}{s_i} = \frac{\mathrm{logit}(y_i/n_i)}{\sqrt{1/y_i+ 1/(n_i-y_i)}}.
\end{equation*}
The selection probability is defined as 
\begin{equation}
P(\mathrm{select} \mid n_i, y_i)=a(t_i)=\Phi(\alpha+\beta t_i),
\label{eq.9}
\end{equation}
where $\beta$ controls the impact of the $t$-statistic on selective publication. We choose the probit function as the selection function according to \cite{copas2013}. Although the Equation~\eqref{nn.5} was derived under the NN model, similar arguments can be made under the setting of the exact distribution for the within-study model. Then the observed distribution of $y_i,n_i$ conditional on published studies is given by
\begin{equation}
  f_O(y_i, n_i) =\mathrm{P}(y_i, n_i \mid \mathrm {select})=\frac{a(t_i) \mathrm{P}(y_i \mid n_i) f_P(n_i)}{\mathrm{P}(\mathrm {select})},
  \label{eq.addx1}
\end{equation}
where the $\mathrm{P(select)}$ denotes the marginal publication probability of the meta-analysis. \textcolor{blue}{Given the $\mathrm{P(select)}$ fixed as $p$, in a similar way to the derivation of~\eqref{nn.3} and~\eqref{nn.5}, we can derive}
\begin{equation}
\begin{aligned}
 f_P(n_i)&=p \frac{1}{\mathrm{P}(\mathrm {select} \mid n_i)} f_O(n_i), \\
 f_O(y_i, n_i) & =\frac{a(t_i) \mathrm{P}(y_i \mid n_i) f_O(n_i)}{\mathrm{P}(\mathrm{select} \mid n_i)},
\end{aligned}
\label{eq.10}
\end{equation}
where the $f_P(\cdot)$ denotes the population probability of $n_i$ among all the studies (published and unpublished) and $f_O(\cdot)$ denotes the conditional probability given published. \textcolor{blue}{Correspondingly to~\eqref{nn.4}, we can derive a constraint,
\begin{equation}
\frac{1}{p} = \frac{1}{N} \sum_{i=1}^N \frac{1}{\mathrm{P}(\mathrm {select} \mid n_i)}.
\label{eq.11}
\end{equation}}
Then, we derive the log-likelihood conditional on published as, 

\begin{equation}
\begin{aligned}
& \ell_O(\Theta)=\sum_{i=1}^N \log f_P(y_i \mid n_i)+\sum_{i=1}^N \log a(t_i)-\sum_{i=1}^N \log \mathrm{P}(\mathrm {select} \mid n_i) +\sum_{i=1}^N \log f_O(n_i),
\end{aligned}
\label{eq.12}
\end{equation}
where $\Theta =(\theta, \tau^2, \beta, \alpha)$, including $\theta ,\tau^2$ in the one-group BN model and $\beta ,\alpha $ in the selection model. We can estimate the bias-adjusted estimator by maximizing the log-likelihood conditional on published with the marginal probability $p$ fixed. To this end, we need to run some iterations. Thus, in each iteration, we need to evaluate each term in~\eqref{eq.12}. \textcolor{blue}{The first term $f_P(y_i\mid n_i)$ corresponds to~\eqref{eq.8}; the second term refers to~\eqref{eq.9}; the last term is free from the parameters, thus can be omitted when estimating parameters through MLE.} The third term $\mathrm{P}(\mathrm{select}\mid n_i)$ is given by 

\begin{equation}
\sum_{j=0}^{n_i} \mathrm{P}(\mathrm {select} \mid n_i, j) \mathrm{P}(j \mid n_i).
\label{eq.13}
\end{equation}

The calculation of $\mathrm{P}(\text{select} \mid n_i, j) \mathrm{P}(j \mid n_i)$ involves an integration:

\begin{equation}
 \mathrm{P}(\mathrm {select} \mid n_i, j) \mathrm{P}(j \mid n_i) = \Phi\left(\alpha+\beta \frac{\log \{j/(n_i-j)\}}{\sqrt{1/j+1/(n_i-j)}}\right) \int_{\theta_i} \tbinom{n_i}{j} \frac{\exp
(\theta _i)^j}{\{1+\exp(\theta_i)\}^{n_i}}\frac 1{\tau }\varphi \left(\frac{\theta
_i-\theta }{\tau }\right)d\theta _i.
\label{eq.14}
\end{equation}
Then, computation of~\eqref{eq.13} in each iteration is demanding and may be infeasible with large $n_i$.
\textcolor{blue}{We solve this computation issue by proposing an approximation method. When $n_i$ gets large, the $t$-statistic $t_i$ asymptotically follows a normal distribution. We can derive the asymptotic expectation and variance for $t_i$ and then fit them to the $\mathrm{P}(\mathrm{select}\mid n_i) = E_{y_i}\{\mathrm{P}(\mathrm{select}\mid n_i, y_i)\} = E_{t_i}\Phi(\alpha+\beta t_i)$ to approximate  $\mathrm{P}(\mathrm{select}\mid n_i)$. We put more details about the approximation method in Web Appendix A and tested the approximation accuracy with different $n_i$ and $p$ in Web Appendix B. We noticed the relative approximation error was less than 1\% in most cases but for small $n_i$ and $p$, the relative error would enlarge to up to 5\%. To avoid the influence of the approximation error on the estimation of parameters, we also developed a program for direct summation. We showed how we structured our code to be more efficient in direct summation and avoid the computational issue in Web Appendix C. We calculated the summation exactly in the analysis of real examples and used the approximation method in simulation studies.}

We estimate the parameters as follows. First, with the constraint function~\eqref{eq.11}, $\alpha$
can be written as a function of the remaining parameters, denoted as $\widehat {\alpha }(\theta ,\tau ,\beta )$ given the fixed $p$.
Then we maximize the log-likelihood conditional on published~\eqref{eq.12} with the other parameters, 
$\ell_O(\Theta )=\ell_O(\theta ,\tau ,\beta ,\widehat{\alpha }(\theta ,\tau ,\beta ))$. The strategy for eliminating $\alpha$ was proposed by \cite{copas2013}. \textcolor{blue}{The MLE of these parameters is denoted by $\widehat{\Theta}=(\widehat{\theta}, \widehat{\tau}, \widehat{\beta})$. We derive the asymptotic variance via the observed inverse fisher information $\boldsymbol{\widehat{\Sigma}}$. Then we derive the approximated 95\% confidence interval (CI) as $[\widehat{\Theta}-1.96\sqrt{\mathrm{diag}(\boldsymbol{\widehat{\Sigma}})},\widehat{\Theta}+1.96\sqrt{\mathrm{diag}(\boldsymbol{\widehat{\Sigma}})}]$.} We then conduct our sensitivity analysis under a series of $p$. 

\section{Extensions to other generalized linear mixed models}
\label{ss:extend}

We consider the meta-analysis with two groups: the treatment and the placebo group. We are interested in estimating the log odds ratio (lnOR) between the two treatments. Following \cite{stijnen2010}, we apply the hypergeometric normal (HN) and the two-group BN model to this case. \textcolor{blue}{As explained in Section~\ref{ss:motiv}, we suppose that the summary statistic of each study, which is used as data in the meta-analysis, was calculated based on individual participant data, which was independently and identically distributed sample from a binary distribution. It ensures the distributional assumptions in this section.}

\subsection{The hypergeometric normal (HN) model}
\label{ss:hn}

Suppose in the $i$-th individual study, the numbers of events in the treatment group and the placebo group are denoted by $y_{{i1}}$ and $y_{{i0}}$ respectively. The total numbers of subjects in the two groups are denoted by $n_{{i1}}$ and $n_{{i0}}$ respectively. The empirical lnOR and its empirical variance for the $i$-th study are given by,

\begin{equation}
\widehat{\theta}_i=\log \left\{\frac{{y_{i1}}/{(n_{i1}-y_{i1})}}{{y_{i0}}/{(n_{i0}-y_{i0})}}\right\}, s_i^2=\frac{1}{y_{i1}}+\frac{1}{n_{i1}-y_{i1}}+\frac{1}{y_{i0}}+\frac{1}{n_{i0}-y_{i0}}.
\label{eq.15}
\end{equation}
It can be used to evaluate the performance of the treatment. The within-study level of the HN model assumes the number of events in the treatment group $y_{i1}$ follows a non-central hypergeometric distribution given the total number of events in the study, $y_i=y_{{i0}}+y_{{i1}}$, fixed. This constraint guarantees the effect size of interest, the common lnOR, is identifiable. The across-study level of the HN model assumes the true lnOR of each study follows a normal distribution, that is $\theta _i\sim N(\theta ,\tau ^2)$, where $\theta$ is the common lnOR across the studies and $\tau$ denotes the heterogeneity parameter. Thus, the population probability mass function of the HN model is given by \cite{stijnen2010} as, 

\begin{equation}
f_P(y_{i1}, y_{i0}\mid n_{i1}, n_{i0}, y_i) = \int \frac{ \tbinom{n_{i1}}{r_{i1}} \tbinom{n_{i0}}{r_{i0}} \exp (\theta_i y_{i1})}{\sum_j\tbinom{n_{i1}}{j} \tbinom{n_{i0}}{y_i-j} \exp (\theta_i j)} \frac{1}{\tau} \varphi\left(\frac{\theta_i-\theta}{\tau}\right) \mathrm{d} \theta_i,
\label{eq.16}
\end{equation}
and the corresponding likelihood function is given by
\begin{equation}
 L(\theta, \tau) =\prod_{i=1}^N f_P(y_{i1}, y_{i0}\mid n_{i1}, n_{i0}, y_i).
 \label{eq.add1}
\end{equation}

The $t$-statistic for the lnOR is derived as the following 

\begin{equation}
t_i=\frac{\widehat{\theta}_i}{s_i}=\frac{\log \{\frac{y_{i1} /(n_{i1}-y_{i1})}{y_{i0} /(n_{i0}-y_{i0})}\}}{\sqrt{\frac{1}{y_{i1}}+\frac{1}{n_{i1}-y_{i1}}+\frac{1}{y_{i0}}+\frac{1}{n_{i0}-y_{i0}}}}.
\label{eq.17}
\end{equation}

Following the same way with the one-group BN model, we can derive the log-likelihood conditional on published taking into account PB as

\begin{equation}
\begin{aligned}
\ell_O(\Theta)&=\sum_{i=1}^N \log f_P(y_{i1}, y_{i0} \mid n_{i1}, n_{i0}, y_i)+\sum_{i=1}^N \log a(t_i)-\sum_{i=1}^N \log \mathrm{P}(\mathrm {select} \mid n_{i1}, n_{i0}, y_i) \\
&+\sum_{i=1}^N \log f_O(n_{i0}, n_{i1}, y_i).
\end{aligned}
\label{eq.18}
\end{equation}

The term $f_P(y_{i1},y_{i0}\mid n_{i1},n_{i0},y_i)$ refers to~\eqref{eq.16}. The term $P(\mathrm{select}\mid n_{i1},n_{i0},y_i)$ is given by

\begin{equation}
\sum_{y_{i1}= \max(0, y_i-n_{i0}) }^{\min(y_i, n_{i1}) } \mathrm{P}(\mathrm {select} \mid n_{i1}, n_{i0}, y_{i1}, y_{i0}) \mathrm{P}(y_{i1}, {y}_{i0} \mid n_{i1}, n_{i0}, y_i).
\label{eq.19}
\end{equation}

We approximate this term through the method using the approximation method in the Web Appendix D. We then follow the
same steps in that for the one-group BN model to estimate the parameters and derive the CI for $\theta$ and $\tau$.

\subsection{The two-group binomial-normal (BN) model}
\label{ss:twogroupbn}

When the total number of events of two groups is relatively small compared with the
group sizes, \cite{bak2018} proved that the HN model can be approximated by a much simpler BN
model with an intercept. Following \cite{stijnen2010}, the within-study model reduced to 

\begin{equation}
y_{i1} \mid \theta_i \sim \mathrm { Binomial }\left(y_i, \frac{\exp \{\log (n_{i1} / n_{i0})+\theta_i\}}{1+\exp \{\log (n_{i1} / n_{i0})+\theta_i\}}\right).
\label{eq.20}
\end{equation}

The population probability mass function for this model is given by \cite{stijnen2010} as

\begin{equation}
f_P(y_{i1},y_{i0}\mid n_{i1},n_{i0},y_i)= \int \tbinom{y_{i}}{y_{i1}} \scalebox{1}{$\frac{\exp \{\log ({n_{i1}}/{n_{i0}})+\theta_i\}^{y_{i1}}}{[1+\exp \{\log ({n_{i1}}/{n_{i0}})+\theta_i\}]^{y_i}} \frac{1}{\tau} \varphi\left(\frac{\theta_i-\theta}{\tau}\right) d \theta_i$},
\label{eq.21}
\end{equation}
and the corresponding likelihood function is given by
\begin{equation}
 L(\theta, \tau)= \prod_{i=1}^N f_P(y_{i1},y_{i0}\mid n_{i1},n_{i0},y_i).
 \label{eq.add2}
\end{equation}

The steps to conduct the sensitivity
analysis are the same with those in the HN model based on~\eqref{eq.17}, ~\eqref{eq.18} and ~\eqref{eq.19}, except for the 
$f_P(y_{i1},y_{i0}\mid n_{i1},n_{i0},y_i)$ is substituted by~\eqref{eq.21} rather than~ \eqref{eq.16}. 

We can also extend our method to the meta-analysis studying the incidence rate with the Poisson normal (PN) model and incidence rate ratios with the BN model (see details in Web Appendix E and F).

\section{Application}
\label{ss:realdata}

We illustrate our methods with the examples that were previously analyzed in \cite{stijnen2010}. We conducted a sensitivity analysis given the marginal selection probability $p$ ranging from 0.1 to 1. When $p=1$, the analysis assumed there was no unpublished studies.

\begin{figure}
 \centering
 \includegraphics{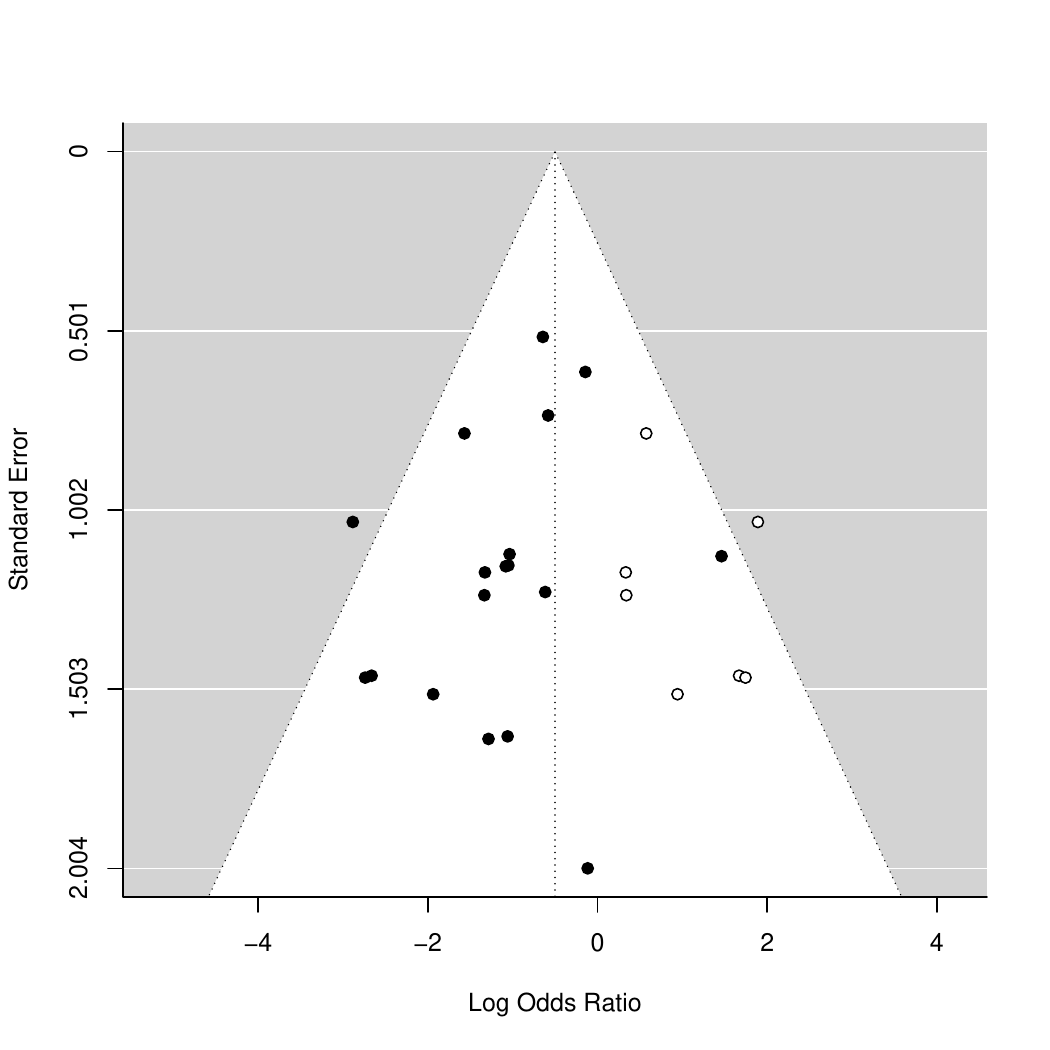}
 \caption{Funnel-plot and trim-and-fill results on the meta-analysis by Neil et al. (2007). The dashed lines are the estimates by trim-and-fill method; the solid dots denote the data in published studies of the meta-analysis; the hollow dots denote the filled studies by trim-and-fill method.}
 \label{fig.2}
\end{figure}

The meta-analysis of 18 clinical trials conducted by \cite{niel2007} evaluated the effect of anti-infective-treated central venous catheters on catheter-related bloodstream infection (CRBSI) in the acute care setting. The data are given in Table~\ref{tab:data1}. A funnel-plot was also given in Figure~\ref{fig.2}, suggesting PB where studies with large risk ratio were less likely to be published. As seen in Table~\ref{tab:data1}, the number of events was low or sometimes zero and sparsity was concerned with. We estimated the overall lnOR between the two treatments using the
HN and two-group BN models. Supposing $p=0.1, 0.2, \cdots, 1$, we evaluated how the estimated lnOR would be impacted by PB. We also conduct the sensitivity analysis under the NN model by \cite{copas2013} for comparisons. The resulting estimates for the lnOR are shown in Table~\ref{tab:resu1}. We noticed that the estimates under the HN
model and the BN model did not have significant differences, which was consistent with the fact that the BN model approximates
the HN model well if the total number of events is small relative to the population of the cohort proved by \cite{bak2018}. However, the results with the NN model and the results with the BN model or the HN model showed a significant difference. It could be inferred that the NN model obtained greatly biased estimates.
On the other hand, with $p$ decreasing from 1 to 0.1, the lnOR increased gradually. When $p$ dropped below 0.7, the significant effect vanished.
It implied that only 8 unpublished studies might eliminate the statistical significant finding, and thus, the result should be interpreted with much caution.
We also observed that with more unpublished studies, the confidence interval was wider.

\begin{table}
 \begin{center}
  \caption{Meta-analysis about the effect of anti-infective-treated central venous catheters on the occurrence of CRBSI by Neil et al. (2007)}
 \label{tab:data1}
 \begin{tabular}{ccccc}
 \hline Study & \multicolumn{2}{c}{ Standard catheter } & \multicolumn{2}{c}{ Anti-infective catheter } \\
 \hline &
 No. of CRBSIs $(y_0)$ &
 No. of patients $(n_0)$ & 
 No. of CRBSIs $(y_1)$ & 
 No. of patients $(n_1)$ \\
 \hline 1 & 3 & 117 & 0 & 116 \\
 2 & 3 & 35 & 1 & 44 \\
 3 & 9 & 195 & 2 & 208 \\
 4 & 7 & 136 & 0 & 130 \\
 5 & 6 & 157 & 5 & 151 \\
 6 & 4 & 139 & 1 & 98 \\
 7 & 3 & 177 & 1 & 174 \\
 8 & 2 & 39 & 1 & 74 \\
 9 & 19 & 103 & 1 & 97 \\
 10 & 2 & 122 & 1 & 113 \\
 11 & 7 & 64 & 0 & 66 \\
 12 & 1 & 58 & 0 & 70 \\
 13 & 5 & 175 & 3 & 188 \\
 14 & 11 & 180 & 6 & 187 \\
 15 & 0 & 105 & 0 & 118 \\
 16 & 1 & 262 & 0 & 252 \\
 17 & 3 & 362 & 1 & 345 \\
 18 & 1 & 69 & 4 & 64 \\
 \hline
 \end{tabular}
 \end{center}
\end{table}

\begin{table}[]
 \begin{center}
  \caption{Estimates of lnOR with the HN model, two-group BN model, and the NN model for meta-analysis by Neil et al. (2007) under $p=1, 0.9, \cdots, 0.1$. \# implies the expected number of unpublished studies. }
 \label{tab:resu1}
 \begin{tabular}{ccccc}
 \hline $p$ & \# & HN & BN & NN \\
 \hline 1.0 & 0 & -1.35(-2.04, -0.67) & -1.30(-1.97, -0.64) & -0.96(-1.42, -0.50) \\ 
 0.9 & 2& -1.21(-1.95, -0.47) & -1.17(-1.88, -0.46) & -0.87(-1.32, -0.42) \\ 
 0.8 & 4& -1.04(-1.87, -0.22) & -1.01(-1.84, -0.19) & -0.78(-1.24, -0.33) \\ 
 0.7 & 8& -0.86(-1.91, 0.20) & -0.84(-1.86, 0.19) & -0.70(-1.17, -0.22) \\ 
 0.6 & 12 & -0.65(-1.98, 0.68) & -0.64(-1.94, 0.66) & -0.62(-1.11, -0.12) \\ 
 0.5 & 18& -0.43(-2.15, 1.27) & -0.43(-2.12, 1.25) & -0.54(-1.07, 0.00) \\ 
 0.4 &27& -0.19(-2.46, 2.08) & -0.21(-2.42, 1.20) & -0.45(-1.04, 0.138) \\ 
 0.3 &42& 0.04(-2.95, 3.04) & 0.01(-2.89, 2.91) & -0.36(-1.02, 0.29) \\ 
 0.2 &72& 0.24(-3.66, 4.13) & 0.20(-3.55, 3.95) & -0.27(-1.02, 0.48) \\ 
 0.1 &162& 0.30(-4.34, 4.94) & 0.28(-4.28, 4.83) & -0.15(-1.07, 0.75) \\ 
 \hline
 \end{tabular}
 \end{center}
\end{table}

Two additional examples in meta-analysis of the log odds of the prevalence and in meta-analysis of the incidence rate ratio are presented in Web Appendix G and H, respectively.

\section{Simulation studies}
\label{ss:simu}

We conducted the simulation studies to substantiate the performance of our proposal.
We considered three settings of experiments. In the Experiments 1-3, we experimented with the one-group BN, the two-group BN and HN models, respectively. 
We set $\theta =-2$ for Experiment 2 and 3 with two-group BN and HN model, but set $\theta=-3$ in Experiment 1 with the one-group BN model to guarantee the sparsity of simulated meta-analysis. We set various heterogeneity parameters, $\tau =0.05, 0.15, 0.3$ to reflect meta-analysis with small, moderate, and relatively large heterogeneity. We generated 1,000 meta-analysis of $S$ studies
including published and unpublished studies. We set $S=15, 25, 50, 100$ to study the performance of our proposal with different scale meta-analysis. 

\textcolor{blue}{In each meta-analysis, we generated $\theta _i$ from $N(\theta ,\tau ^2)$. Then, in Experiment 1, we sampled $n_i$ from a discrete uniform distribution $U(200,400)$, and $y_i\sim \mathrm{Binomial}\left(n_i, \exp(\theta_i)/\left(1+\exp(\theta_i)\right)\right)$ according to~\eqref{eq.7}. In Experiment 2 and 3, we sampled $n_{i1}$ and $n_{i0}$ separately from a discrete uniform distribution $U(200,400)$, and $y_i$ from a discreate uniform distribution $U(15, 25)$. We then generate $y_{i1}$ from the binomial distribution~\eqref{eq.20} in Experiment 2 and the non-central hypergeometric distribution~\eqref{eq.16} in Experiment 3.}

For each meta-analysis dataset, we fixed the marginal selection probability $p=0.6$, suggesting approximately 60\% of studies will be published. We set $\beta$ in the selection function~\eqref{eq.9} as 2 and an appropriate $\alpha$ was sought for. 
Then we decided whether a study was published with a Bernoulli distribution, that is $\mathrm{Bern}(\Phi(\alpha+\beta t_i))$, where $t_i$ corresponded to~\eqref{eq.9} for the one-group BN model and~\eqref{eq.14} for the two-group BN and HM models, respectively. The average events rate for each simulated dataset were summarized in Table~\ref{tab:simu1} to show their sparsity. We could see the average events rate is less than 5\% in Experiment 1 and less than 1\% for Experiment 2 and 3, suggesting that datasets of meta-analysis with sparse data were successfully generated.

\begin{table}[]
\caption{Average event rates (\%) for the simulated datasets. (Full datasets include both the published and unpublished studies; the published datasets only include published studies. For two-group situations, we reported the smaller average event rate of the treatment and the placebo group.)}
 \label{tab:simu1}
 \centering
 \resizebox{\linewidth}{!}{

 \begin{tabular}{ccccccccccc}
\hline \multirow{2}{*}{ Experiment } & \multirow{2}{*}{ Model } & \multirow{2}{*}{ $\tau$ } & \multicolumn{2}{c}{$S=15$} & \multicolumn{2}{c}{$S=25$} & \multicolumn{2}{c}{$S=50$} & \multicolumn{2}{c}{$S=100$} \\
\cline{4-11} & & & Full & Published & Full & Published & Full & Published & Full & Published \\
\hline \multirow{3}{*}{1} & \multirow{3}{*}{ One-group BN } & 0.05 & 4.76 & 4.76 & 4.75 & 4.75 & 4.75 & 4.75 & 4.75 & 4.75 \\
 & & 0.15 & 4.8 & 4.8 & 4.78 & 4.78 & 4.8 & 4.8 & 4.79 & 4.79 \\
 & & 0.3 & 4.93 & 4.93 & 4.94 & 4.94 & 4.92 & 4.92 & 4.93 & 4.93 \\
\hline \multirow{3}{*}{2} & \multirow{3}{*}{ Two-group BN } & 0.05 & 0.82 & 0.82 & 0.82 & 0.82 & 0.82 & 0.82 & 0.81 & 0.81 \\
 & & 0.15 & 0.83 & 0.83 & 0.82 & 0.82 & 0.82 & 0.82 & 0.82 & 0.82 \\
 & & 0.3 & 0.85 & 0.85 & 0.84 & 0.84 & 0.84 & 0.84 & 0.84 & 0.84 \\
\hline \multirow{3}{*}{3} & \multirow{3}{*}{$\mathrm{HN}$} & 0.05 & 0.85 & 0.85 & 0.85 & 0.85 & 0.86 & 0.86 & 0.86 & 0.86 \\
 & & 0.15 & 0.86 & 0.86 & 0.86 & 0.86 & 0.86 & 0.86 & 0.86 & 0.86 \\
 & & 0.3 & 0.87 & 0.87 & 0.88 & 0.88 & 0.88 & 0.88 & 0.88 & 0.88 \\
\hline
\end{tabular}
 }
\end{table}

We performed our proposed method only with published studies. The mean values of estimates for $\theta$, the sample standard deviation across 1,000 simulated datasets for $\theta$, and the 95\% CI coverage probability for the true value of $\theta$ were summarized in Table~\ref{tab:simu2}. We also compared our proposed method with the sensitivity analysis method based on the NN model by \cite{copas2013} and MLE for the GLMM only with published studies. From the simulation results, we found that our proposal obtained coverage probabilities close to the nominal level and less bias compared with the other two methods. The NN model by \cite{copas2013} performed the worst with the lowest coverage probabilities, indicating the inadequate capability of the NN model to address PB in sparse data. We also found a decreased coverage probability when the true heterogeneity $\tau$ increased, showing that the true value of $\tau$ would affect the estimation of $\theta$. With large heterogeneity across studies of meta-analysis, estimations of the random-effects models might fail to cover the true value due to large data variation. \textcolor{blue}{We also examined the performance of our proposal and the other two methods in estimating $\tau$ (see Web Appendix I). We noticed that no methods based on MLE including the MLE using both published and unpublished studies can well estimate the heterogeneity parameter. Under small heterogeneity $\tau=0.05$, all methods obtain unsatisfactory coverage probabilities far lower than the nominal level.}

\begin{sidewaystable}
 \caption{Simulation results for estimations of $\theta$. (MLE(published): MLE with only published studies; AVE: mean values of estimates over all simulations; SD: sample standard deviation over all simulations; CP: 95\% Confidence interval coverage probability) }
 \label{tab:simu2}
 \resizebox{\linewidth}{!}{
 
 \begin{tabular}{ccccccccccccc}
\hline \multirow{2}{*}{ Experiment } & \multirow{2}{*}{ Model } & \multirow{2}{*}{ $\tau$ } & \multirow{2}{*}{ Method } & \multirow{2}{*}{ True value } & \multicolumn{2}{c}{$S=15$} & \multicolumn{2}{c}{$S=25$} & \multicolumn{2}{c}{$S=50$} & \multicolumn{2}{c}{$S=100$} \\
\cline{6-13} & & & & & AVE(SD) & CP & AVE(SD) & CP & AVE(SD) & CP & AVE(SD) & CP \\
\hline \multirow{9}{*}{1} & \multirow{9}{*}{ One-group BN } & \multirow{3}{*}{0.05} & Proposed & \multirow{9}{*}{-3} & $-3.04(0.13)$ & 0.88 & $-3.03(0.1)$ & 0.89 & $-3.01(0.06)$ & 0.92 & $-3.01(0.04)$ & 0.9 \\
 & & & \cite{copas2013} & & $-3.08(0.12)$ & 0.7 & $-3.08(0.09)$ & 0.67 & $-3.08(0.07)$ & 0.59 & $-3.08(0.05)$ & 0.49 \\
 & & & MLE(published) & & $-3.06(0.1)$ & 0.85 & $-3.06(0.08)$ & 0.8 & $-3.06(0.06)$ & 0.73 & $-3.06(0.04)$ & 0.59 \\
\cline{6-13} & & \multirow{3}{*}{0.15} & Proposed & & $-3.04(0.14)$ & 0.87 & $-3.03(0.1)$ & 0.87 & $-3.01(0.07)$ & 0.88 & $-3.01(0.05)$ & 0.88 \\
 & & & \cite{copas2013} & & $-3.08(0.12)$ & 0.72 & $-3.08(0.1)$ & 0.72 & $-3.08(0.07)$ & 0.67 & $-3.08(0.05)$ & 0.52 \\
 & & & MLE(published) & & $-3.07(0.12)$ & 0.82 & $-3.07(0.09)$ & 0.79 & $-3.07(0.07)$ & 0.72 & $-3.07(0.05)$ & 0.54 \\
\cline{6-13} & & \multirow{3}{*}{0.3} & Proposed & & $-3.05(0.18)$ & 0.84 & $-3.03(0.13)$ & 0.85 & $-3.01(0.09)$ & 0.88 & $-3.01(0.06)$ & 0.92 \\
 & & & \cite{copas2013} & & $-3.08(0.16)$ & 0.77 & $-3.08(0.12)$ & 0.8 & $-3.09(0.09)$ & 0.75 & $-3.08(0.06)$ & 0.63 \\
 & & & MLE(published) & & $-3.1(0.16)$ & 0.77 & $-3.1(0.12)$ & 0.75 & $-3.11(0.09)$ & 0.62 & $-3.11(0.06)$ & 0.42 \\
\hline \multirow{9}{*}{2} & \multirow{9}{*}{ Two-group BN } & \multirow{3}{*}{0.05} & Proposed & \multirow{9}{*}{-2} & $-2.04(0.22)$ & 0.93 & $-2.03(0.15)$ & 0.94 & $-2.03(0.11)$ & 0.94 & $-2.02(0.07)$ & 0.94 \\
 & & & \cite{copas2013} & & $-1.97(0.22)$ & 0.66 & $-1.98(0.18)$ & 0.68 & $-2(0.13)$ & 0.72 & $-2.01(0.09)$ & 0.76 \\
 & & & MLE(published) & & $-1.94(0.3)$ & 0.91 & $-1.93(0.22)$ & 0.9 & $-1.92(0.16)$ & 0.86 & $-1.92(0.12)$ & 0.8 \\
\cline{6-13} & & \multirow{6}{*}{0.15} & Proposed & & $-2.03(0.22)$ & 0.93 & $-2.03(0.15)$ & 0.95 & $-2.02(0.11)$ & 0.94 & $-2.02(0.08)$ & 0.95 \\
 & & & \cite{copas2013} & & $-1.97(0.23)$ & 0.65 & $-1.97(0.18)$ & 0.66 & $-1.99(0.13)$ & 0.71 & $-2(0.09)$ & 0.77 \\
 & & & MLE(published) & & $-1.94(0.32)$ & 0.9 & $-1.92(0.23)$ & 0.88 & $-1.92(0.16)$ & 0.85 & $-1.91(0.12)$ & 0.8 \\
\cline{6-13} & & & Proposed & & $-2.01(0.24)$ & 0.9 & $-2(0.17)$ & 0.92 & $-2(0.12)$ & 0.92 & $-2(0.08)$ & 0.94 \\
 & & & \cite{copas2013} & & $-1.93(0.25)$ & 0.61 & $-1.94(0.18)$ & 0.65 & $-1.96(0.14)$ & 0.68 & $-1.97(0.1)$ & 0.71 \\
 & & & MLE(published) & & $-1.91(0.33)$ & 0.86 & $-1.89(0.25)$ & 0.86 & $-1.89(0.18)$ & 0.81 & $-1.89(0.13)$ & 0.76 \\
\hline \multirow{9}{*}{3} & \multirow{9}{*}{$\mathrm{HN}$} & \multirow{3}{*}{0.05} & Proposed & \multirow{9}{*}{-2} & $-2.03(0.23)$ & 0.92 & $-2.02(0.15)$ & 0.96 & $-2.01(0.11)$ & 0.94 & $-2.01(0.07)$ & 0.96 \\
 & & & \cite{copas2013} & & $-1.94(0.23)$ & 0.7 & $-1.95(0.17)$ & 0.71 & $-1.96(0.13)$ & 0.74 & $-1.96(0.09)$ & 0.72 \\
 & & & MLE(published) & & $-1.93(0.31)$ & 0.9 & $-1.91(0.21)$ & 0.89 & $-1.9(0.16)$ & 0.82 & $-1.89(0.11)$ & 0.75 \\
\cline{6-13} & & \multirow{3}{*}{0.15} & Proposed & & $-2.02(0.23)$ & 0.92 & $-2.01(0.16)$ & 0.94 & $-2(0.11)$ & 0.94 & $-2(0.07)$ & 0.95 \\
 & & & \cite{copas2013} & & $-1.93(0.23)$ & 0.71 & $-1.93(0.18)$ & 0.69 & $-1.94(0.13)$ & 0.71 & $-1.95(0.09)$ & 0.73 \\
 & & & MLE(published) & & $-1.93(0.31)$ & 0.89 & $-1.91(0.23)$ & 0.87 & $-1.89(0.16)$ & 0.82 & $-1.88(0.11)$ & 0.75 \\
\cline{6-13} & & \multirow{3}{*}{0.3} & Proposed & & $-2(0.24)$ & 0.91 & $-1.99(0.16)$ & 0.94 & $-1.99(0.12)$ & 0.92 & $-1.98(0.08)$ & 0.93 \\
 & & & \cite{copas2013} & & $-1.89(0.23)$ & 0.66 & $-1.9(0.18)$ & 0.64 & $-1.92(0.13)$ & 0.68 & $-1.92(0.1)$ & 0.63 \\
 & & & MLE(published) & & $-1.9(0.33)$ & 0.87 & $-1.87(0.24)$ & 0.84 & $-1.87(0.17)$ & 0.8 & $-1.86(0.12)$ & 0.69 \\
\hline
\end{tabular} } 
\end{sidewaystable}

\textcolor{blue}{We conducted additional simulation studies with the PN and BN for meta-analysis of incidence rate ratios and placed the simulation results in Web Appendix J. We found that our proposal also outperformed the other two approaches. 
In summary, our model can well adjust for PB in evaluating the effect size of the meta-analysis with several GLMMs.}

\section{Discussion}
\label{ss:discuss}

Our model serves as the first solution to the issue of PB under the GLMM with the exact within-study model. Compared with the NN model, the GLMMs with the exact within-study model are less likely to be influenced by the sparse data. We extend these GLMMs with a sensitivity analysis to debias them from potential PB with the Copas $t$-statistic selection model. Based on the Copas $t$-statistics selection model, \cite{huang2023} developed a method to adjust for PB using the information on unpublished studies obtained from clinical trial registries. Their method also relies on the NN model. In sparse data, GLMMs are more appealing. \textcolor{blue}{ The Copas $t$-statistic selection model is more interpretable compared with the Copas-Heckman selection model since it bridges the selection mechanism for published studies with their $p$-values. However, it is hard to identify the true selective mechanism in reality. \cite{zhou2024copasheckmantype} proposed a sensitivity analysis with GLMMs based on Copas-Heckman selection model. We recommend conducting comprehensive sensitivity analysis with multiple selection functions in practice.}

Though our method showed good performance with both real-data analysis and simulation studies, there also
remain some limitations. First, there is still a small gap between the estimated effect size and the
true effect size especially in small-scale meta-analysis with large across-study heterogeneity. A potential cause is that our approximation method suffers from small approximation errors (see Web Appendix B). It is appealing to develop a more accurate approximation method to reduce this bias. \textcolor{blue}{Secondly, our proposal as well as other methods based on MLE performs poorly in estimating the heterogeneity parameter since MLE is a biased estimator for $\tau$ with finite samples. If researchers are interested in the heterogeneity parameters or some statistics related to them, our method may fail to give satisfactory solutions. \cite{dersimonian1986}, \cite{dersimian2007} and \cite{higgins2002} studied more efficient methods compared with MLE to estimate the heterogeneity parameter. These methods were mostly based on the NN model and were hard to fit into our methods. It is important to adjust our method to better estimate the heterogeneity parameter in further work.}

\section*{Acknowledgements}
This research was partly supported by Grant-in-Aid for Challenging Exploratory Research (16K12403) and for Scientific Research (16H06299, 18H03208) from the Ministry of Education, Science, Sports and Technology of Japan.

\section*{Supplementary Materials}
Web Appendices, Tables, and Figures, and data and code referenced in Sections~\ref{ss:propose} to~\ref{ss:discuss} are available with this paper at the Biometrics website on Oxford Academic.

\section*{Data Availability}
\label{software}

The data used in this paper are available for download in Supplementary Materials.
\vspace*{-8pt}

\bibliographystyle{biom} 
\bibliography{biomtemplate}

\label{lastpage}

\end{document}